\documentclass[hyper,12pt]{JHEP3}

\newcommand{\Frac}[2]{\frac{\displaystyle #1}{\displaystyle #2}}
\newcommand{\beq}{\begin{equation}}
\newcommand{\eeq}{\end{equation}}
\newcommand{\beqn}{\begin{eqnarray}}
\newcommand{\eeqn}{\end{eqnarray}}
\newcommand{\beqns}{\begin{eqnarray*}}
\newcommand{\eeqns}{\end{eqnarray*}}
\title{ Electromagnetic penguin operators and direct CP
violation in $K\rightarrow\pi \ell^+\ell^-$ }
\author{Giancarlo D'Ambrosio
\\Theory Division, CERN, CH-1211 Geneva 23,
 Switzerland\\
{\rm and}\\
INFN- Sezione
di Napoli, I-80126 Naples, Italy\\
\email{E-mail: Giancarlo.D'Ambrosio@cern.ch}}
\author{Dao-Neng Gao\\Department of
Astronomy and Applied Physics, University of Science and
Technology of China, Hefei, Anhui, 230026, China \\
\email{E-mail: gaodn@ustc.edu.cn}}

\received{}
\accepted{}
\abstract{ Supersymmetric extensions of
the Standard Model predict a large enhancement of the Wilson
coefficients of the dimension-five electromagnetic penguin
operators affecting the direct CP violation in
$K_L\rightarrow\pi^0 e^+ e^-$ and the charge asymmetry in
$K^\pm\rightarrow\pi^\pm \ell^+ \ell^-$.
 Here we compute the relevant matrix elements in
the chiral quark model and compare these with the ones given by
lattice calculations.}


\preprint{CERN-TH/2002-057\\INFNNA-IV-2001/26\\March  2002}

\begin{document}

\section{Introduction}

Rare kaon decays  provide an ideal place both to test the
Standard Model (SM) and to unravel new physics beyond it
\cite{DI98}. The origin of CP violation is still an open question
in modern particle physics.  Dimension-five operators including
the electromagnetic and chromomagnetic penguin operators (EMO and
CMO) play important roles in these studies since the CP-violating
effects from these operators are suppressed in the SM but could be
enhanced in its extensions \cite{BCIRS99,CIP99, DIM99,
HMPV99,BLMM00}. In fact present experiments, HyperCP
\cite{HyperCP} and KLOE \cite{KLOE}, and planned ones, NA48b
\cite{NA48b}, are going to  substantially improve the present
limits on the Wilson coefficients of these operators by studying
CP-violating asymmetries in $K^\pm \rightarrow 3 \pi$, $K^\pm
\rightarrow \pi \pi \gamma$ and in $K^\pm \rightarrow \pi^\pm \ell
\bar{\ell}$ ($\ell=e,~\mu$). As we shall see, although it is hard
to test the SM now it is possible to probe interesting new physics
scenarios. To this purpose it is necessary to know hadronic matrix
elements accurately: we address this issue in a particular
bosonization scheme.

The weak effective Hamiltonian, contributed by EMO and CMO, can be written
as \cite{BCIRS99, BLMM00}
\beq\label{HMO}
{\cal H}_{\rm eff}=C_\gamma^+(\mu) Q_\gamma^+(\mu)+
C_\gamma^-(\mu) Q_\gamma^-(\mu)+C_g^+(\mu) Q_g^+(\mu)
+C_g^-(\mu) Q_g^-(\mu)+ {\rm h.c.},\label{cpg}
\eeq
where $C^\pm_{\gamma,~g}$ are the Wilson coefficients and
\beqn
Q_\gamma^\pm=\Frac{e Q_d}{16\pi^2}(\bar{s}_L\sigma_{\mu\nu}d_R\pm
\bar{s}_R\sigma_{\mu\nu}d_L)F^{\mu\nu},\label{EMO}\\
Q_g^\pm=\Frac{g}{16\pi^2}(\bar{s}_L\sigma_{\mu\nu}t_a d_R\pm
\bar{s}_R\sigma_{\mu\nu}t_a d_L) G_a^{\mu\nu}.\label{CMO} \eeqn
Here $\sigma_{\mu\nu}=i/2[\gamma_\mu, \gamma_\nu]$. The SM
structure, $SU(2)_L \times U(1)$, imposes
 a chiral suppression for the following operators
\cite{RPS93, BME94}:
 \beqn {\cal
H}^{\rm SM}_{\rm eff}=
\Frac{G_F}{\sqrt{2}}V_{td}V^*_{ts}\left[C_{11}\frac{g}{8\pi^2}
(m_d\bar{s}_L\sigma_{\mu\nu}t_a
d_R+m_s\bar{s}_R\sigma_{\mu\nu}t_a d_L)G_a^{\mu\nu}\nonumber \right.\\
\left.+C_{12}\frac{e}{8\pi^2}(m_d\bar{s}_L\sigma_{\mu\nu}d_R+m_s\bar{s}_R
\sigma_{\mu\nu}d_L)F^{\mu\nu}\right]+ {\rm h.c.}, \label{SMHMO}
\eeqn
and
\beqn
C_{11}(m_W)=\Frac{3x^2}{2(1-x)^4}{\rm
ln}~x-\Frac{x^3-5x^2-2x}{4(1-x)^3},\label{SMC11}\\
C_{12}(m_W)=\Frac{x^2(2-3x)}{2(1-x)^4}{\rm
ln}~x-\Frac{8x^3+5x^2-7x}{12(1-x)^3}, \label{SMC12} \eeqn where
$x=m_t^2/m_W^2$ and $t_a$ are the $SU(3)$-matrices.
However, as we shall see,
new flavour structures in the supersymmetry-breaking
terms allow us to avoid the chiral suppression for the operators in
eq. (\ref{SMHMO}).

Among rare kaon decays, the flavour-changing neutral current
(FCNC) transitions $K\rightarrow\pi \ell^+ \ell^-$, induced at the
one-loop level in the SM, are well suited to explore its quantum
structure and extensions \cite{EPR87, DEIN95,DI98}. The decay
$K_L\rightarrow \pi^0 e^+ e^-$ receives contributions from three
sources \cite{DI98,EPR88,SE88}: direct CP violation,
 indirect CP violation  due to $K^0$--$\bar{K}^0$ mixing,
and CP conservation from the two-photon rescattering in
$K_L\rightarrow \pi^0\gamma\gamma$. Therefore, once long-distance
effects have been carefully disentangled \cite{SE88}, new physics,
induced by the operators in  eq. (\ref{EMO}), can be probed in
this channel. Analogously the charge asymmetry in $K^\pm\rightarrow
\pi^\pm \ell^+ \ell^-$ could be enhanced by a large Wilson coefficient
of the operator in  eq. (\ref{EMO}) \cite{ME02}.
Recently it has been shown that also  
T-odd correlations in charged K$_{l4}$-decays depend upon the 
effective Hamiltonian in (\ref{cpg}) \cite{Retico}.

We thus  consider here the matrix element $\langle
\pi^0|Q_\gamma^+|K^0\rangle$ to determine  the observables
discussed above. In order to evaluate the
bosonization of the EMO  we exploit the chiral quark model, which
provides an effective link between QCD and low energy chiral
perturbation theory. This is particularly interesting since the
first lattice calculation of the matrix element $\langle
\pi^0|Q_\gamma^+|K^0\rangle$ has been done in Ref. \cite{BLMM00}
and thus a comparison of the two methods can be performed. This
might be useful in general to understand the extent of
validity of the two
approaches in  the evaluation of other matrix elements such as the
penguin operator.

\section{The chiral quark model}

The chiral quark model ($\chi$QM ) \cite{MG84} has been
extensively used to study  low energy hadronic physics
involving strong and weak interactions \cite{ERT90,PR91,
BEF94, ABEF96, AET98,MKG99}.
Note that the interactions among
mesons proceeds in this model  only by means of quark loops:
starting from the short-distance effective Hamiltonian in terms of
quark operators (such as four-quark operators, EMO, and CMO), the
$\chi$QM allows us to deduce the low energy
effective lagrangian in terms
 of the input parameters of the model.

In the $\chi$QM \cite{ERT90}, a term that  represents the coupling
between the light (constituent) quarks and the Goldstone mesons
\beq \label{QMmass} -M_Q(\bar{q}_R U q_L+\bar{q}_L U^+ q_R) \eeq
has been introduced into the QCD lagrangian.  The Goldstone meson
fields, $\phi(x)$, are collected in a unitary $3\times 3$ matrix $
U={\rm exp} (i/f_\pi \lambda\cdot \phi (x))$ (where the $\lambda ^a $'s are the
$3\times3$ Gell-Mann matrices and $f_\pi\simeq 93$ MeV) with
det$U$=1, which transforms as \beq U\rightarrow V_R U V_L^+ \eeq
under chiral SU(3)$_L\times$SU(3)$_R$ transformations ($V_L$,
$V_R$), and \beqn\label{Phi}
\frac{1}{\sqrt{2}}\lambda\cdot\phi(x)=\left(\begin{array}{ccc}
\frac{\pi^0}{\sqrt{2}}+\frac{\eta_8}{\sqrt{6}} &\; \pi^+ &\;K^+ \\ \; \\
\pi^- &\;-\frac{\pi^0}{\sqrt{2}}+\frac{\eta_8}{\sqrt{6}} &\; K^0\\ \; \\
K^- &\;\bar{K}^0 & \; -\frac{2\eta_8}{\sqrt{6}}\\
\end{array}\right) \ . \eeqn

In the presence of the term (\ref{QMmass}), it is convenient to use new
quark fields, $Q_L$ and $Q_R$, called  ``rotated basis", defined as follows
\beqn
&&Q_L=\xi q_L,~~~~~~~~~\bar{Q}_L=\bar{q}_L\xi^+,\nonumber \\
&&Q_R=\xi^+ q_R,~~~~~~~\bar{Q}_R=\bar{q}_R\xi,\label{basis} \eeqn
with $\xi$ chosen such that \beq U=\xi^2. \eeq
The chiral
SU(3)$_L\times$SU(3)$_R$ transformation \beq \xi(x)\rightarrow
V_R\xi(x) h^+(x)=h(x)\xi(x)V_L^+ \eeq  defines the
compensating SU(3)$_V$ transformation $h(\phi(x))$,
which is the wanted
ingredient for a non-linear representation of the chiral group.
Then $Q_{L,~R}$'s transform as \beq Q_L\rightarrow h(x)
Q_L,~~~~~~~Q_R\rightarrow h(x) Q_R, \eeq while the term
(\ref{QMmass}) \beq -M_Q(\bar{q}_R U q_L+\bar{q}_L U^+ q_R)
=-M_Q(\bar{Q}_R Q_L+\bar{Q}_L Q_R) \eeq is invariant. Therefore,
the quark fields $Q_{L,~R}$ can be interpreted as ``constituent
chiral quarks" and $M_Q$ as a ``constituent quark mass".

Now in order to evaluate the bosonization of the EMO, we firstly
write down the EMO using the ``rotated basis" in the Euclidean
space \beqn\label{EMOr} {\cal H}^{\rm SM}_{\rm
eff}&=&\bar{Q}\left(\frac{1-\gamma_5}{2}\xi^+\lambda \xi^+
m_s+\frac{1+\gamma_5}{2}\xi\lambda\xi m_d\right)\sigma_{\mu\nu}Q
~C_{\rm EMO}
F^{\mu\nu}\nonumber\\
&&+\bar{Q}\left(\frac{1+\gamma_5}{2}\xi\lambda^+\xi
m_s+\frac{1-\gamma_5}{2}\xi^+\lambda^+\xi^+
m_d\right)\sigma_{\mu\nu}Q~C^*_{\rm EMO}F^{\mu\nu}, \eeqn where
$\lambda_{ij}=\delta_{i3}\delta_{j2}$, and \beq C_{\rm
EMO}=\frac{G_F}{\sqrt{2}}\frac{e}{8\pi^2}\lambda_t~
C_{12},~~~~~\lambda_t=V_{td}V^*_{ts}. \eeq Here we use the form of
EMO in the SM [eq. (\ref{SMHMO})]. It is very easy to extend it to
the general form in eq. (\ref{HMO}).

Then the effective action induced by the EMO can be written as
follows \beq\label{action32} \Gamma_E({\cal A},M)=-\frac{1}{2}\int
d^4 x~{\rm Tr}~\int_0^\infty\frac{d\tau}{\tau}\int\frac{d^d
p_E}{(2\pi)^d} {\rm exp}\left[{-\tau(p_E^2+M^2_Q)}\right]~{\rm
exp}({-\tau {\cal D}^\prime}),\eeq where Tr is the trace over
colour, flavour and Lorentz space, ${\cal D}^{\prime}$ is defined in
(\ref{dp}), and the integral over $\tau$ is introduced by using
the  proper time method \cite{Ball89}. The detailed
derivation for eq. (\ref{action32}) has been shown in the Appendix,
and  dimensional regularization has been used for the
involved divergences. Expanding ${\rm exp}({-\tau {\cal
D}^\prime})$ in powers of $\tau$, and integrating over the
momenta, one can get the effective action in powers of $\tau$, and
the corresponding coefficients are the so-called Seeley--DeWitt
coefficients.  Then the effective lagrangian can be obtained  by
integrating out $\tau$.
The standard procedure can be found in
Refs. \cite{Ball89, ERT90}. If we set $F_1=F_2=J_{\mu\nu}=0$ in
${\cal D}^\prime$ [see (\ref{dp}) in Appendix], which implies
that the EMO is switched off, eq. (\ref{action32}) will give the
same effective lagrangian as in Ref.
\cite{ERT90}. Here we are concerned about the effective lagrangian
generated from the EMO, which is relevant to $K\rightarrow \pi
\ell^+ \ell^-$ transitions. Thus at the leading order we get
\beq\label{LEMOSM} {\cal L}^{\rm SM}_{\rm EMO}=\Frac{iN_C
M_Q}{8\pi^2}C_{\rm EMO}\langle m_d\lambda U L_\mu L_\nu+m_s
\lambda  L_\mu L_\nu U^+\rangle F^{\mu\nu}+ {\rm h.c.}, \eeq where
$L_\mu=iU^+D_\mu U$, $N_C$ is the number of  colours, and
$\langle A\rangle$ denotes the trace of $A$ in the flavour space.
Likewise, the corresponding effective lagrangian from the general
form of the EMO in eq. (\ref{HMO}) is \beq\label{LEMO} {\cal
L}^\pm_{\rm EMO}=\Frac{iN_CM_Q}{8\pi^2}
\Frac{eQ_d}{16\pi^2}C_\gamma^\pm \langle \lambda U L_\mu
L_\nu\pm\lambda L_\mu L_\nu U^+\rangle F^{\mu\nu}+ {\rm h.c.} \eeq
where  ${\cal L}^+  _{\rm EMO}$ (${\cal L}^-  _{\rm EMO}$)
generates parity-even (odd) transitions.

The matrix elements of the EMO between a $K^0$ and a $\pi^0$
can be written as \beq \langle
\pi^0|Q_\gamma^+|K^0\rangle=i\Frac{\sqrt{2}e Q_d}{16\pi^2
m_K}p_\pi^\mu p_K^\nu F_{\mu\nu}B_T, \label{kpg1} \eeq \beq
\langle \pi^0|Q_\gamma^-|K^0\rangle=0. \eeq
Then from eqs. (\ref{HMO})
and (\ref{LEMO}), we can obtain \beq\label{BT} B_T=\frac{N_C M_Q
m_K}{4\pi^2 f_\pi^2}. \eeq Setting $M_Q=$0.3~GeV, we have
$B_T=1$.31, which is consistent with $B_T=1.18\pm0.09$ found in
the lattice
\cite{BLMM00} and  $B_T\simeq 1$ in Ref. \cite{RPS93},
and the range $|B_T|=0.5\sim2$ adopted in Ref. \cite{BCIRS99}.
Our theoretical error on  $B_T$ in (\ref{BT}) has two sources: 
i) from the quark mass 
$M_Q$, which we believe it is very small, $\sim 10 \% $, and ii) from 
higher order corrections in the $\chi$QM, generated by large-$N_c$ gluonic 
interactions. We have evaluated this contribution using the standard 
techniques in 
Refs. \cite{ERT90,PR91,deraf95}, finding the  correction
 to (\ref{BT}) 
\beq\label{gluon}\frac{\pi ^2}{9N_c} \frac{\langle \frac{\alpha _s}{\pi} GG
\rangle}{M_Q^4}.\eeq
The size of the gluon condensate cannot be simply related to the one
which appears in the QCD sum rule \cite{deraf95}. 
However terms like the one
in (\ref{gluon}),  but with larger coefficients, correct also the leading order
predictions for the $L_i$'s and  $f_\pi$
\cite{ERT90}. 
Model consistency and the phenomenologically successful predictions of 
 the  leading order evaluation, lead us to the reasonable expectation 
that the  gluon correction
in (\ref{gluon}) cannot exceed $\sim 30 \% $ and so consequently
 we can very conservatively 
estimate the error in this way on  $B_T$, i.e. $B_T=1.31\pm 0.4$.

We stress that the agreement with the lattice is found for natural values
of the chiral quark model. So we can be quite confident in this result.

\section{$K\rightarrow\pi \ell^+ \ell^-$}

The  decay width of $K_L\rightarrow\pi^0 e^+ e^-$ induced by the EMO
is given by
\beq\label{br1} {\rm Br}(K_L\rightarrow\pi^0 e^+
e^-)_{\rm EMO}=8.9\times10^3 ~{\rm GeV^2}~B_T^2~|{\rm Im}
C_\gamma^+|^2.
\eeq 
To obtain an interesting  bound on ${\rm Im} C_\gamma^+$ we improve our error
on  $B_T$ by considering also the lattice results  \cite{BLMM00}.
Thus from the experimental upper bound
\cite{KTeV00} \beq\label{exp1} {\rm Br}(K_L\rightarrow\pi^0 e^+
e^-)< 5.1\times10^{-10}, \eeq  
we get
\beq\label{IMCG} |{\rm Im} C_\gamma^+|< 1.8\times 10^{-7} ~{\rm
GeV^{-1}} \eeq
at $80\%$ C.L.

\vspace{0.4cm}

It is known that $K^\pm\rightarrow\pi^\pm \ell^+ \ell^-$ is
dominated by long-distance, charge-symmetric, one-photon exchange
\cite{EPR87,SW90,LWS92,DEIP98}. This piece can be written as
\cite{DEIP98} \beq\label{LDKP} A(K^+\rightarrow\pi^+ \ell^+
\ell^-)=-\Frac{e^2}{m_K^2(4\pi)^2} W_+(z) (p_K+p_\pi)^\mu
\bar{u}(p_-)\gamma_\mu v(p_+), \eeq where $z=(p_K-p_\pi)^2/m_K^2$,
and the general form factor $W_+(z)$ has been shown in  Ref.
\cite{DEIP98}. 
The piece induced by the EMO will interfere 
with
the imaginary part of
$W_+(z)$, which arises from the two-pion intermediate state
 \cite{DEIP98}. The
 asymmetry is then written as \beq\label{def}
\left(\frac{\delta\Gamma}{2\Gamma}\right)^{\rm
EMO}_{\ell}=\Frac{|\Gamma(K^+\rightarrow\pi^+ \ell^+
\ell^-)-\Gamma(K^-\rightarrow\pi^- \ell^+\ell^-)|_{\rm
EMO}}{\Gamma(K^+\rightarrow\pi^+
\ell^+\ell^-)+\Gamma(K^-\rightarrow\pi^- \ell^+\ell^-)}. \eeq
Interestingly, with a kinematical
cut $z\ge 4m_{\pi}^2/m_K^2$, the charge asymmetry in eq. (\ref{def})
could be substantially enhanced \cite{ME02}. Thus from eqs.
(\ref{LDKP}) and (\ref{def}), and using the upper
bound of $|{\rm Im} C_\gamma^+|$ given in eq. (\ref{IMCG}), we can
find the charge asymmetry for $\ell=e,~\mu$ as \beq\label{nocut}
\left(\frac{\delta\Gamma}{2\Gamma}\right)^{\rm EMO}_e< 1.3\times
10^{-4},\;\;\; \;\left(\frac{\delta\Gamma}{2\Gamma}\right)^{\rm
EMO}_\mu< 4.5\times 10^{-4}
 \eeq without the kinematical cut for $z$, and
 \beq\label{cut}
 \left(\frac{\delta\Gamma}{2\Gamma}\right)^{\rm EMO}_e< 1.2\times
10^{-3},\;\;\;\;\left(\frac{\delta\Gamma}{2\Gamma}\right)^{\rm
EMO}_\mu< 1.3\times 10^{-3}
 \eeq
 with the cut $z\ge 4 m_{\pi}^2/m_K^2$. Note that, differently from  Ref.
 \cite{ME02}, here we only use the experimental bound of
 Br($K_L\to\pi^0 e^+ e^-$)
to estimate the charge asymmetry in both electron and muon
mode. So we are neglecting possible lepton-family violations.

\section{Limits on new flavour structures}

From eq. (\ref{LEMOSM}), one can get ${\rm Im}
C_\gamma^+$ in the SM \beq\label{CGSM} |{\rm Im} C_\gamma^+|^{\rm
SM}=\frac{3 G_F}{\sqrt{2}}(m_s+m_d)|{\rm Im} \lambda_t~C_{12}|.
\eeq Due to the smallness of ${\rm Im}\lambda_t\sim 10^{-4}$, this
contribution from the SM is strongly suppressed, and far smaller
than the upper bound (\ref{IMCG}). Therefore in the following we
turn our attention to physics beyond the SM.

Among the possible new physics scenarios, low energy supersymmetry
(SUSY) \cite{SUSY}, represents one of the most interesting and
consistent extensions of the SM. In generic supersymmetric models,
the large number of new particles carrying flavour quantum numbers
would naturally lead to large effects in CP violation and FCNC
amplitudes \cite{FCNC}. Particularly, one can generate the
enhancement of $C^\pm_{\gamma,~g}$ at one-loop, via intermediate
squarks and gluinos, which is due both to the strong coupling
constant and to the removal of chirality suppression present in
the SM. Full expressions for the Wilson coefficients generated by
gluino exchange at the SUSY scale can be found in Ref.
\cite{GGMS96}. We are interested here only in the contributions
proportional to $m_{\tilde{g}}$, which are given by \beqn
C_{\gamma,{\rm
SUSY}}^\pm(m_{\tilde{g}})=\Frac{\pi\alpha_s(m_{\tilde{g}})}{m_{\tilde{g}}}
\left[(\delta^{\rm D}_{\rm LR})_{21}\pm(\delta^{\rm D}_{\rm
LR})^*_{12}\right]
F_{\rm SUSY}(x_{gq})\label{CEMO},\\
C_{g,{\rm
SUSY}}^\pm(m_{\tilde{g}})=\Frac{\pi\alpha_s(m_{\tilde{g}})}{m_{\tilde{g}}}
\left[(\delta^{\rm D}_{\rm LR})_{21}\pm(\delta^{\rm D}_{\rm
LR})^*_{12}\right] G_{\rm SUSY}(x_{gq})\label{CCMO}, \eeqn where
$(\delta^{\rm D}_{\rm LR})_{ij}=(M^2_{\rm D})_{i_{\rm L}j_{\rm
R}}/m^2_{\tilde{g}}$ denotes the off-diagonal entries of the
(down-type) squark mass matrix in the super-CKM basis,
$x_{gq}=m^2_{\tilde{g}}/m^2_{\tilde{q}}$ with $m_{\tilde{g}}$
being the average gluino mass and $m_{\tilde{q}}$ the average
squark mass. The explicit expressions of $F_{\rm SUSY}(x)$ and
$G_{\rm SUSY}(x)$ are given in Ref. \cite{BCIRS99}, but noting
that they do not depend strongly on $x$,
 it is sufficient,  for our purposes, to
approximate $F_{\rm SUSY}(x)\sim$ $F_{\rm SUSY}(1)=2/9$ and
$G_{\rm SUSY}(x)\sim$ $G_{\rm SUSY}(1)=-5/18$.
In any case it  will be easy to extend the numerology once
$x_{gq}$ is better known.
 Also the determination of the Wilson coefficients in eqs. (\ref{CEMO}) and
(\ref{CCMO}) can be improved by the renormalization group analysis
\cite{BCIRS99, BLMM00}. 
Then by taking $m_{\tilde{g}}=500$
GeV, $m_t=174$ GeV,
$m_b=5$ GeV, and $\mu=m_c=1.25$ GeV, we will have
\beq
\left|{\rm Im} C_\gamma^+\right|^{\rm SUSY}=2.4\times 10^{-4}{\rm
GeV^{-1}}~\left|{\rm
Im} [(\delta^{\rm D}_{\rm LR})_{21}+(\delta^{\rm D}_{\rm
LR})^*_{12}]\right|.
\eeq
From eq. (\ref{IMCG}), we obtain
\beq
\left|{\rm Im} [(\delta^{\rm D}_{\rm LR})_{21}+(\delta^{\rm D}_{\rm
LR})^*_{12}]\right|<7.7\times 10^{-4},
\eeq
comparable with the one given by the lattice calculation \cite{BLMM00}.

\section{Conclusions}

To conclude, supersymmetric extensions of the SM may enhance
the Wilson coefficients of the electromagnetic penguin operators.
This leads to  interesting phenomenology to be studied:
  the direct CP violation
in $K_L\rightarrow \pi^0 e^+ e^-$ and the charge asymmetry in
$K^\pm\rightarrow\pi^\pm \ell^+ \ell^-$. To this purpose we
evaluate the relevant matrix element in the $\chi$QM.
Interestingly we find a very good agreement with lattice results
for the natural parameters of the model \cite{BLMM00}.
 The present experimental upper bound of ${\rm
Br}(K_L\rightarrow\pi^0 e^+ e^-)$ allows to obtain an  upper bound of
$|{\rm Im} C_\gamma^+|$, and thus to predict the upper bound of the
charge asymmetry in $K^\pm\rightarrow\pi^\pm \ell^+ \ell^-$
induced by EMO. The analysis shows that the predictions for the
relevant matrix elements are solid and thus high precision
measurements of CP-observables might probe interesting extensions
of the SM.


\acknowledgments

We thank Gino Isidori for useful discussions. D.N.G. is supported
in part by the NSF of China under Grant No. 19905008.

\appendix
\newcounter{pla}
\renewcommand{\thesection}{\Alph{pla}}
\renewcommand{\theequation}{\Alph{pla}.\arabic{equation}}
\setcounter{pla}{1} \setcounter{equation}{0}

\section{Appendix}

Here we present the derivation for eq. (\ref{action32}) in the
$\chi$QM. Including the constituent quark mass term in eq.
(\ref{QMmass}), the strong lagrangian in the rotated basis
[eq.(\ref{basis})] and in the Euclidean space is (after we switch
off contributions from the EMO) \beq\label{QCDE} {\cal L}^{E}_{\rm
Str}=-\frac{1}{4}G^a_{\mu\nu}G^a_{\mu\nu}+\bar{Q}D_{E} Q, \eeq
where $G^a_{\mu\nu}$ is the gluon fields strength tensor, and
$D_E$ the Euclidean Dirac operator \beq D_E={\gamma}_\mu
\nabla_\mu+M={\gamma}_\mu(\partial_\mu+{\cal A}_\mu)+M, \eeq with
\beq {\cal A}_\mu=iG_\mu+\Gamma_\mu-\frac{i}{2}\gamma_5\xi_\mu,
~~~~M=-\frac{1}{2}(\Sigma-\gamma_5 \Delta)-M_Q. \eeq
Note that, in
the present paper, we use the same notations as in Ref. \cite{PR91} and so
for the Euclidean quantities, $\gamma_\mu^+=\gamma_\mu$,
$\{\gamma_\mu, \gamma_\nu\}=2\delta_{\mu\nu}$, and
$\sigma_{\mu\nu}=-i/2[\gamma_\mu, \gamma_\nu]$.
 The external
vector and axial-vector fields now  appear in $\Gamma_\mu$ and
$\xi_\mu$ \beqn
\Gamma_\mu=\frac{1}{2}[\xi^+(\partial_\mu-i~r_\mu)\xi+
\xi(\partial_\mu-i~l_\mu)\xi^+],\\
\xi_\mu=i[\xi^+(\partial_\mu-i~r_\mu)\xi-\xi(\partial_\mu-i~l_\mu)\xi^+],
\eeqn and \beq \Sigma=\xi^+{\cal M}\xi^++\xi{\cal
M}\xi,~~~~\Delta=\xi^+{\cal M}\xi^+-\xi{\cal M}\xi. \eeq Here
${\cal M}$ is the current quark mass matrix, and \beq
\Gamma_\mu^+=-\Gamma_\mu,~~~\xi_\mu^+=\xi_\mu,~~~\Sigma^+=\Sigma,
~~~\Delta^+=-\Delta,~~~{\cal M}^+={\cal M}. \eeq
 The $\Sigma$- and
$\Delta$-terms break chiral symmetry explicitly.

The Euclidean effective action $W_E(U,r,l,{\cal M},M_Q)$ is
obtained as follows \beq {\rm exp} W_E(U,r,l,{\cal
M},M_Q)=\frac{1}{Z}\int{\cal D} G_\mu {\rm
exp}\left( -\frac{1}{4}G^a_{\mu\nu}G^a_{\mu\nu}\right) {\rm exp}
\Gamma_E({\cal A},M), \eeq where $Z$ is the normalization factor,
and \beq {\rm exp} \Gamma_E({\cal A},M)=\int{\cal D}\bar{Q}{\cal
D} Q {\rm exp}\int d^4 x \bar{Q}D_E Q={\rm det} D_E. \eeq Since we
are concerned with the non-anomalous part of the effective action,
we have \beq \label{action0} \Gamma_E({\cal A},M)=\frac{1}{2}{\rm
ln~det} D^+_E  D_E, \eeq with \beq
D_E^+=-\gamma_\mu \left(\partial_\mu+i
G_\mu+\Gamma_\mu+\frac{i}{2}\gamma_5\xi_\mu\right)
-\frac{1}{2}(\Sigma+\gamma_5 \Delta)-M_Q. \eeq
Using the technique of the heat kernel expansion \cite{Ball89},
one can derive the
 effective strong lagrangian starting from (\ref{action0}), which has
been discussed extensively in the literature.

Now we switch on the EMO. Note that this operator has been
expressed using the rotated basis in eq. (\ref{EMOr}); it is thus
easy to know that (\ref{action0}) should become
 \beq\label{action1}
\Gamma_E({\cal A},M)=\frac{1}{2}{\rm ln~det} {D^+_E}^\prime
{D_E}^\prime, \eeq with \beq {D_E}^\prime=D_E+J,\;\;\;
{D^+_E}^\prime=D^+_E+J^+, \eeq \beq J=\sigma_{\mu\nu}J_{\mu\nu},
\;\;\;J^+=\sigma_{\mu\nu}J^+_{\mu\nu}, \eeq and \beqn
J_{\mu\nu}&=&-\left(\frac{1-\gamma_5}{2}\xi^+\lambda\xi^+m_s+\frac{1+
\gamma_5}{2}\xi\lambda\xi m_d\right)C_{\rm EMO} F_{\mu\nu}\nonumber\\
&&-\left(\frac{1+\gamma_5}{2}\xi\lambda^+\xi
m_s+\frac{1-\gamma_5}{2}\xi^+\lambda^+\xi^+ m_d\right)C^*_{\rm
EMO}F_{\mu\nu},\\ \nonumber\\
J_{\mu\nu}^+&=&J_{\mu\nu}~(\gamma_5~\leftrightarrow - \gamma_5).
\eeqn Thus, one can get \beq\label{DDE} {D^+_E}^\prime
{D_E}^\prime-M_Q^2=-\nabla_\mu\nabla_\mu+E+F_1+F_2, \eeq with \beq
E=iM_Q\gamma_\mu\gamma_5\xi_\mu-\frac{i}{2}\sigma_{\mu\nu}R_{\mu\nu},
\eeq \beq F_1=-\gamma_\mu\sigma_{\alpha\beta}d_\mu
J_{\alpha\beta}+\frac{i}{2}\gamma_\mu\sigma_{\alpha\beta}\{
\gamma_5\xi_\mu,J_{\alpha\beta}\}-M_Q\sigma_{\mu\nu}(J_{\mu\nu}+J^+_{\mu\nu}),
\eeq \beq F_2=-4i\gamma_\mu J_{\mu\nu}\nabla_\nu, \eeq and \beq
R_{\mu\nu}=iG_{\mu\nu}-i\left(\frac{1+\gamma_5}{2}\xi^+F_{R\mu\nu}\xi
+\frac{1-\gamma_5}{2}\xi F_{L\mu\nu}\xi^+\right). \eeq
Here we set
$\Sigma=\Delta=0$, $d_\mu$  is the covariant derivative with
respect to the $\Gamma_\mu$-connection, i.e. $d_\mu
A=\partial_\mu A+[\Gamma_\mu, A]$, and the relation \beq
[\gamma_\mu,\sigma_{\alpha\beta}]=2i(\delta_{\mu\alpha}\gamma_\beta-
\delta_{\mu\beta}\gamma_\alpha) \eeq has been used. We only
include the linear terms of $J_{\mu\nu}$ in $F_1$ and $F_2$
because we are concerned about the $O(G_F)$ $\Delta S=1$
transitions.

Starting from (\ref{action1}), and in terms of the proper
time method \cite{Ball89}, we have \beq\label{action2}
\Gamma_E({\cal A},M)=-\frac{1}{2}\int d^4 x~{\rm
Tr}~\int_0^\infty\frac{d\tau}{\tau}\langle x|{\rm exp}(-\tau
{D^+_E}^\prime {D_E}^\prime)|x\rangle, \eeq where the trace is
taken in colour, flavour, and Lorentz space.  By inserting a
complete set of plane waves and using (\ref{DDE}), we obtain
\beq\label{action3} \Gamma_E({\cal A},M)=-\frac{1}{2}\int d^4
x~{\rm Tr}~\int_0^\infty\frac{d\tau}{\tau}\int\frac{d^d
p_E}{(2\pi)^d} {\rm exp}\left[{-\tau(p_E^2+M^2_Q)}\right]~{\rm
exp}({-\tau {\cal D}^\prime}), \eeq where \beq\label{dp} {\cal
D}^\prime=E-\nabla\cdot\nabla+F_1-2 i p_E\cdot
\nabla+F_2+4\gamma_\mu {p_E}_\nu J_{\mu\nu}. \eeq


\vspace{0.4cm}
\begin{thebibliography}{40}
\bibitem{DI98}G. D'Ambrosio and G. Isidori, Int. J. Mod. Phys. A {\bf 13}
(1998) 1 [arXiv:hep-ph/9611284], and references therein.
\bibitem{BCIRS99}A. J. Buras, G. Colangelo, G. Isidori, A. Romanino, and
L. Silvestrini, Nucl. Phys. B {\bf 566} (2000) 3
[arXiv:hep-ph/9908371].
\bibitem{CIP99}G. Colangelo, G. Isidori, and J. Portol\'es, Phys. Lett. B
{\bf 470} (1999) 134 [arXiv:hep-ph/9908415].
\bibitem{DIM99}G. D'Ambrosio, G. Isidori, and G. Martinelli, Phys. Lett. B
{\bf 480} (2000) 164 [arXiv:hep-ph/9911522].
\bibitem{HMPV99}X.-G. He, H. Murayama, S. Pakvasa, and G. Valencia,
Phys. Rev. D {\bf 61} (2000) 071701 [arXiv:hep-ph/9909562].
\bibitem{BLMM00}D. Becirevic, V. Lubicz, G. Martinelli, and F. Mescia,
Phys. Lett. B {\bf 501} (2001) 98 [arXiv:hep-ph/0010349].
\bibitem{HyperCP}H. K. Park {\it et al.}, HyperCP Collaboration,
Phys. Rev. Lett. {\bf 88} (2002) 111801 [arXiv:hep-ex/0110033].
\bibitem{NA48b}M. Calvetti, in the Proceedings of the
International Conference on CP violation (KAON2001), Pisa, 2001;
 http://www1.cern.ch/NA48.
\bibitem{KLOE} T. Spadaro, KLOE Collaboration, in the Proceedings
of the International Conference on CP violation (KAON2001), Pisa,
2001.
\bibitem{RPS93}Riazuddin, N. Paver, and F. Simeoni,
Phys. Lett. B {\bf 316} (1993) 397 [arXiv:hep-ph/9308328].
\bibitem{BME94}
N.G. Deshpande, Xiao-Gang He and S. Pakvasa, Phys. Lett. B {\bf326}
(1994) 307 [arXiv:hep-ph/9401330]; S. Bertolini, M. Fabbrichesi,
and E. Gabrielli, Phys. Lett. B {\bf 324} (1994) 164
[arXiv:hep-ph/9311293].
\bibitem{EPR87}G. Ecker, A. Pich, and E. de Rafael, Nucl. Phys. B {\bf
291} (1987) 692.
\bibitem{DEIN95}G. D'Ambrosio, G. Ecker, G. Isidori, and H. Neufeld, {\it
Radiative non-leptonic kaon decays}, in {\it The Second DA$\phi$NE Physics
Handbooks}, eds. L. Maiani, G. Pancheri, and \ N. Paver, LNF (1995).
\bibitem{EPR88}G. Ecker, A. Pich, and E. de Rafael, Nucl. Phys. B {\bf
303} (1988) 665.
\bibitem{SE88}L.M. Sehgal, Phys. Rev\emph{.} D \textbf{38} (1988) 808;
P.\ Heiliger and L.M. Sehgal, Phys. Rev\emph{.} D \textbf{47}
 (1993) 4920;
 L. Cappiello, G. D'Ambrosio and M. Miragliuolo, Phys. Lett. B
\textbf{298} (1993) 423; A. G. Cohen, G. Ecker and \ A. Pich, Phys.
Lett. B \textbf{304} (1993) 347;  G. Ecker, A. Pich and E. de
Rafael, Phys. Lett. B \textbf{237} (1990) 481; G. D'Ambrosio and
J. Portol{\'e}s, Nucl. Phys. B
 \textbf{492} (1997) 417 [arXiv:hep-ph/9610244]; F.~Gabbiani and G.~Valencia,
Phys.\ Rev. D {\bf 64} (2001) 094008 [arXiv:hep-ph/0105006].
\bibitem{ME02}A. Messina, Phys. Lett. B {\bf 538} (2002)
130 [arXiv:hep-ph/0202228].
\bibitem{Retico}
A.~Retico,
Phys.\ Rev.\ D {\bf 65} (2002) 117901 
[arXiv:hep-ph/0203044].
\bibitem{MG84}A. V. Manohar and H. Georgi, Nucl. Phys. B {\bf 234}
(1984) 189.
\bibitem{ERT90}D. Espriu, E. de Rafael, and J. Taron, Nucl. Phys. B {\bf
345} (1990) 22; Erratum, {\it ibid.} B {\bf 355} (1991) 278.
\bibitem{PR91}A. Pich and E. de Rafael, Nucl. Phys. B {\bf 358}
(1991) 311.
\bibitem{BEF94}S. Bertolini, J.O. Eeg, and M. Fabbrichesi,
Nucl. Phys. B {\bf 449} (1995) 197 [arXiv:hep-ph/9409437].
\bibitem{ABEF96}V. Antonelli, S. Bertolini, J.O. Eeg, and M. Fabbrichesi,
Nucl. Phys. B {\bf 469} (1996) 143 [arXiv:hep-ph/9511255].
\bibitem{AET98}A.A. Andrianov, D. Espriu, and R. Tarrach,
Nucl. Phys. B {\bf 533} (1998) 429 [arXiv:hep-ph/9803232].
\bibitem{MKG99}M. Franz, H.-C. Kim, and K. Goeke, Nucl. Phys. B {\bf
562} (1999) 213 [arXiv:hep-ph/9903275].
\bibitem{Ball89}R. D. Ball, Phys. Rep. {\bf 161} (1989) 1.
\bibitem{deraf95} E.~de Rafael,
``Chiral Lagrangians and kaon CP violation,''
[arXiv:hep-ph/9502254].
\bibitem{KTeV00}KTeV Collaboration, A. Alavi-Harati {\it et al.},
Phys. Rev. Lett. {\bf 86} (2001) 397 [arXiv:hep-ex/0009030].
\bibitem{SW90}M. J. Savage and M. B. Wise, Phys. Lett. B {\bf 250}
(1990) 151.
\bibitem{LWS92}M. Lu, M. B. Wise, and M. J. Savage, Phys. Rev. D {\bf 46}
(1992) 5026 [arXiv:hep-ph/9207222].
\bibitem{DEIP98}G. D'Ambrosio, G. Ecker, G. Isidori,
and J. Portol\'es, JHEP {\bf 08} (1998) 004
[arXiv:hep-ph/9808289].
\bibitem{SUSY}H. P. Nilles, Phys. Rep. {\bf 110} (1984) 1;
H. Haber and G. Kane, Phys. Rep. {\bf 117} (1985) 75.
\bibitem{FCNC}S. Dimopoulos and H. Georgi, Nucl. Phys. B {\bf  193}
(1981) 150; J. Ellis and D.V. Nanopoulos, Phys. Lett. B {\bf 110}
(1982) 44; R. Barbieri and R. Gatto, Phys. Lett. B {\bf 110}
(1982) 211; M. J. Duncan, Nucl. Phys. B {\bf 221} (1983) 285; J. F.
Donoghue, H. P. Nilles, and \ D. Wyler, Phys. Lett. B {\bf 128}
(1983) 55; L. J. Hall, V.A. Kostelecky, and S. Raby, Nucl. Phys. B
{\bf 267} (1986) 415.
\bibitem{GGMS96}F. Gabbiani, E. Gabrielli, A. Masiero, and L. Silvestrini,
Nucl. Phys. B {\bf 477} (1996) 321 [arXiv:hep-ph/9604387].
\end{thebibliography}
\end{document}